\documentclass[11pt]{article}
\pdfoutput=1 
\usepackage{jheppub}
\usepackage{hyperref}
\usepackage{booktabs}
\usepackage{cleveref}
\usepackage[T1]{fontenc} 
\usepackage{siunitx} 
\usepackage{xcolor,graphicx}
\usepackage{dcolumn}
\usepackage{bm}
\usepackage{braket} 
\usepackage{amsmath}
\usepackage{slashed}
\usepackage{MnSymbol}
\usepackage{fontawesome}

\usepackage{bbold}

\newcommand{\cL}{\mathcal{L}}

\counterwithin*{equation}{section}
\allowdisplaybreaks

\newcommand{\nnl}{\nonumber \\}

\newcommand{\cM}{{\mathcal M}}

\newcommand{\dslash}[1]{#1 \! \! \! {\bf /}}

\renewcommand{\vec}[1]{{\mathbf{#1}}}

\def\hc{{\rm h.c.}}

\robustify\bfseries

\begin{document}

\title{Unleashing the Power of EFT \\ in Neutrino--Nucleus Scattering} 

\author[a,b]{Joachim Kopp,}
\author[c]{Noemi Rocco,}
\author[d]{Zahra Tabrizi,}

\affiliation[a]{Theoretical Physics Department, CERN, Geneva, Switzerland}
\affiliation[b]{PRISMA Cluster of Excellence \& Mainz Institute for Theoretical Physics, \\
                Johannes Gutenberg University, Staudingerweg 7, 55099 Mainz, Germany}
\affiliation[c]{Theoretical Physics Department, Fermi National Accelerator Laboratory, \\
                P.O. Box 500, Batavia, IL 60510, USA}
\affiliation[d]{Northwestern University, Department of Physics \& Astronomy, \\
                2145 Sheridan Road, Evanston, IL 60208, USA}

\emailAdd{jkopp@cern.ch}
\emailAdd{nrocco@fnal.gov}
\emailAdd{ztabrizi@northwestern.edu}
          
\abstract{\,Neutrino physics is advancing into a precision era with the construction of new experiments, particularly in the few GeV energy range. Within this energy range, neutrinos exhibit diverse interactions with nucleons and nuclei. This study delves in particular into neutrino--nucleus quasi-elastic cross sections, taking into account both standard and, for the first time, non-standard interactions, all within the framework of effective field theory (EFT). The main uncertainties in these cross sections stem from uncertainties in the nucleon-level form factors, and from the approximations necessary to solve the nuclear many-body problem. We explore how these uncertainties influence the potential of neutrino experiments to probe new physics introduced by left-handed, right-handed, scalar, pseudoscalar, and tensor interactions. For some of these interactions the cross section is enhanced, making long-baseline experiments an excellent place to search for them. Our results, including tabulated cross sections for all interaction types and all neutrino flavors, can serve as the foundation for such searches. \href{https://github.com/ztabrizi/EFT-in-Neutrino-Nucleus-Scattering/}{\faGithub}}

\maketitle

\section{Introduction}
\label{sec:Introduction}

How do neutrinos interact with matter? While the answer to this question is well understood at the qualitative level, making quantitative predictions for neutrino--nucleus cross sections has proven to be a daunting task and is a leading source of systematic uncertainty in present and upcoming precision neutrino experiments.  It may therefore seem foolish to even try to complicate matters further by considering the possible impact of physics beyond the Standard Model (SM). In this paper, we will nevertheless take a step forward in precisely this direction. Working in the context of effective field theory (EFT), we will compute neutrino--nucleus interaction cross sections including new couplings with arbitrary Lorentz structure.  These results can form the cornerstone of future experimental searches for ``new physics'' in the neutrino sector, as well as for phenomenological work relating extensions of the SM to observables.

Our starting point is SM Effective Field Theory (SMEFT) \cite{Buchmuller:1985jz, Grzadkowski:2010es}, which is defined by requiring the same gauge symmetries and the same particle content as in the SM, but adding non-renormalizable operators to parameterize the low-energy footprint of particles much heavier than the electroweak scale. But while SMEFT is a suitable framework for new physics searches at the electroweak scale -- that is, at the LHC -- for lower-energy probes such as hadron/meson decay and accelerator-based neutrino experiments, it is more appropriate to work with a different EFT, one in which electroweak-scale degrees of freedom ($W$, $Z$, Higgs, top) have been integrated out.  This theory is called Weak Effective Field Theory (WEFT) or Low-Energy Effective Field Theory (LEFT) in the literature. It can be ultra-violet (UV) completed by SMEFT, and the two theories can be related to one another by renormalization group running and parameter matching at the electroweak scale \cite{%
  Cirigliano:2012ab,      
  Falkowski:2017pss,      
  Jenkins:2017jig,        
  Jenkins:2017dyc,        
  Falkowski:2018dmy,      
  Aebischer:2018bkb,      
  Descotes-Genon:2018foz, 
  Bischer:2019ttk,        
  Dekens:2019ept,         
  Falkowski:2019kfn,      
  Falkowski:2020pma,      
  Falkowski:2021vdg,      
  Falkowski:2021bkq}.     
(This is of course assuming that there are no new particles at the electroweak scale or below. If new physics exists already at a scale lower than the $W$ boson mass, WEFT would be consistent up to the scale of the new physics, and it would need to be UV-completed by an extension of SMEFT that includes the new light degrees of freedom.) At even lower energy scales, namely in the regime of quasi-elastic (QE) scattering, the relevant hadronic degrees of freedom are nucleons instead of quarks. QE scattering makes a significant and often dominant contribution to the total detection rate in current and future accelerator-based long-baseline experiments with GeV-scale beams.

Calculating neutrino cross-sections in the QE regime requires first describing the interactions of the neutrino with individual nucleons, and then taking into account nuclear effects. 
Accurate predictions at low and moderate neutrino energies have been achieved through ``ab-initio'' methods like Green's Function Monte Carlo and the Coupled Cluster approaches~\cite{Lovato:2020kba, Lovato:2017cux, Lovato:2016gkq, Sobczyk:2023sxh}. For higher neutrino energies, relevant for accelerator neutrino experiments, alternative approaches utilize the factorization of the nuclear final state. The spectral function (SF) formalism~\cite{Benhar:1994hw, Benhar:2006wy, Benhar:2010nx}, adopted in this work, incorporates relativistic effects in both kinematics and the interaction vertex, while providing a precise depiction of nuclear dynamics. While the findings presented here are specifically focused on the QE region, the SF approach has been generalized to describe multiple reaction mechanisms. Previous works~\cite{Rocco:2018mwt, Rocco:2020jlx} demonstrate its application to phenomena like meson exchange currents and pion production. Additionally, alternative methods such as SuSA-v2 and the relativistic mean-field approach also provide predictions of neutrino-nucleus cross sections up to high neutrino energies, encompassing a diverse array of reaction mechanisms~\cite{Amaro:2019zos,Gonzalez-Jimenez:2014eqa, Megias:2016lke, Amaro:2021sec}.

The calculation of neutrino cross sections in the QE regime for arbitrary WEFT interactions and including nuclear effects, as well as the discussion of the associated uncertainties, are the main topics of this paper. The most important novel challenge we face is that nucleon-level form factors have to be precisely known for arbitrary WEFT inteactions. 

In the past, physics beyond the SM in the neutrino sector has often been discussed in the framework of ``non-standard interactions'', which is similar to WEFT: the relevant degrees of freedom are those present below the electroweak scale (but above the QCD scale), and the new operators are dimension-six four-fermion interactions \cite{%
    Wolfenstein:1977ue, 
    Valle:1987gv,       
    Grossman:1995wx,    
    Kopp:2007mi,        
    Kopp:2007ne,        
    Antusch:2008tz,     
    Gavela:2008ra,      
    Ohlsson:2012kf,     
    Miranda:2015dra,    
    deGouvea:2015ndi,   
    Coloma:2023ixt}.    
However, this widely used formalism has several shortcomings (see e.g.\ the discussion in~\cite{Falkowski:2019kfn}): (i) it typically does not take into account renormalization group effects and the necessity of an embedding into a theory that respects electroweak symmetry. This means that correlations between operators that would arise from such an embedding are typically neglected. (ii) even more problematic, some authors treat operators affecting neutrino production and those affecting neutrino detection separately, even though in many cases, they are the same. First steps in resolving these shortcomings have been taken in refs.~\cite{Falkowski:2019xoe, Falkowski:2019kfn}, which introduce a comprehensive EFT framework for neutrino oscillations, offering a systematic approach applicable to experiments with extended baselines and diverse neutrino production and detection processes. The practicality of this formalism has been demonstrated in ref.~\cite{Falkowski:2021bkq} by applying it to the FASER$\nu$ neutrino experiment at the LHC, which showcases the capability of FASER$\nu$ to probe new physics scales up to $\sim \SI{10}{TeV}$ for certain operators.

In the following, we will first briefly introduce the effective field theory that we work in (\cref{subsec:WEFT}) and then proceed to the calculation of first the neutrino--nucleon interaction amplitudes for all EFT operators (\cref{sec:nu-nucleon}), and then the interaction cross sections including nuclear effects (\cref{sec:nu-nucleus}). We will present our results in \cref{sec:XSections} and discuss their uncertainties, before summarizing and concluding in \cref{sec:Conclusion}.

Our main cross-section results in tabulated form are available from \href{https://github.com/ztabrizi/EFT-in-Neutrino-Nucleus-Scattering/}{GitHub} \cite{github}.

\section{Formalism}
\label{sec:Formalism}

\subsection{Weak Effective Field Theory}
\label{subsec:WEFT}

In this section we introduce the effective field theory (EFT) formalism in which we will calculate neutrino--nucleus cross section in the presence of physics beyond the SM. In particular, we are interested in Weak Effective Field Theory (WEFT), valid below the electroweak scale, with the electroweak gauge bosons, the Higgs boson, and the top quark integrated out. The part of the WEFT Lagrangian that is relevant to neutrino--nucleus interactions is given by~\cite{Falkowski:2021bkq}:
\begin{align} 
    \cL_{\rm WEFT} 
    &\supset
    - \,\frac{2 V_{ud}}{v^2} \Big\{
      [ {\bf 1} + \epsilon_L]_{\alpha\beta}
             (\bar{q}_u \gamma^\mu P_L q_d) (\bar\ell_\alpha \gamma_\mu P_L \nu_\beta)
    \,+\, [\epsilon_R]_{\alpha\beta} (\bar{q}_u \gamma^\mu P_R q_d)
                                          (\bar\ell_\alpha \gamma_\mu P_L \nu_\beta) \nnl
    &\quad +\,
      \frac{1}{2} [\epsilon_S]_{\alpha\beta}
             (\bar{q}_u q_d) (\bar\ell_\alpha P_L \nu_\beta)
    - \frac{1}{2} [\epsilon_P]_{\alpha\beta}
             (\bar{q}_u \gamma_5 q_d) (\bar\ell_\alpha P_L \nu_\beta) \nnl 
    &\quad +\,
      \frac{1}{4} [\epsilon_T]_{\alpha\beta} (\bar{q}_u \sigma^{\mu\nu} P_L q_d)
                                      (\bar\ell_\alpha \sigma_{\mu\nu} P_L \nu_\beta)
    + \hc \Big \} \,.
  \label{eq:EFT_lweft}
\end{align}
Here, $v \equiv (\sqrt{2} G_F)^{-1/2} \approx \SI{246}{GeV}$ is the vacuum expectation value (vev) of the Higgs field, $V_{ud}$ is the $ud$ element of the Cabibbo--Kobayashi--Maskawa (CKM) matrix, $P_{L,R} = \frac{1}{2} (1 \mp \gamma^5)$ are chirality projection operators, and $\sigma^{\mu\nu} = \tfrac{i}{2} [\gamma^\mu,\gamma^\nu]$. The $q_u$ and $q_d$ fields describe up and down quarks, respectively.  The charged leptons are labeled $\ell_\alpha$ (with $\alpha$ their flavor index), and the neutrino fields in the flavor basis are $\nu_\alpha$. They can be connected to the mass eigenstates $\nu_k$ using the Pontecorvo--Maki--Nakagawa--Sakata (PMNS) mixing matrix: $\nu_{\alpha} = \sum_{k=1}^3 U_{\alpha k} \nu_k$. The first term of \cref{eq:EFT_lweft} describes the SM charged current (CC) interaction between neutrinos, charged leptons and quarks, while the subsequent terms add the new left-handed ($L$), right-handed ($R$), scalar ($S$), pseudo-scalar ($P$) and tensor ($T$) interactions to the Lagrangian. These terms are dimension-six, and their dimensionless Wilson coefficients $[\epsilon_X]_{\alpha\beta}$ (with $X=L,R,S,P,T$) describe the interaction strengths of the new operators relative to SM weak interactions. 

It is worth emphasizing that the SM parameters entering this Lagrangian, namely $v$ and $V_{ud}$ can also be impacted by new physics. For instance, modifications to the decay rate of muons will affect the measurement of the Fermi constant, $G_F$, from which $v$ is derived. Similarly, new physics in nuclear beta decays can affect the extraction of $V_{ud}$. As discussed in the literature \cite{Bhattacharya:2011qm, Gonzalez-Alonso:2016etj, Falkowski:2017pss, Descotes-Genon:2018foz}, these possible biases can be reabsorbed into a redefinition of $\epsilon_L$. We give these redefinitions explicitly in \cref{sec:WEFT-redefinition}. Experimental data analyzed with $v$ and $V_{ud}$ set to the measured values as reported for instance by the Particle Data Group (PDG) \cite{Workman:2022ynf} would therefore constrain this redefined $\epsilon_L$. In the following, $\epsilon_L$ will be understood in that sense.

\subsection{Neutrino--Nucleon Scattering}
\label{sec:nu-nucleon}

We are interested in studying the effects that the new interactions introduced in the Lagrangian of \cref{eq:EFT_lweft} have on charged current quasi-elastic (CCQE) neutrino interactions. At the nucleon level, the interactions we are interested in are 
\begin{align}
  \nu_\beta + n \to \ell_\alpha^- + p^+ \,
\end{align}
for neutrinos, and 
\begin{align}
  \bar\nu_\beta+ p^+ \to \ell_\alpha^+ + n \,
\end{align}
for anti-neutrinos.The structure of the WEFT Lagrangian implies that the amplitudes related to the neutrino--nucleon interactions take the form
\begin{align}
  \cM_{\alpha\beta}  &= \delta_{\alpha\beta} A_{L,\alpha}
    + \sum_{X=L,R,S,P,T} [\epsilon_X]_{\alpha \beta} A_{X,\alpha} \,,
  \label{eq:Mdecomposition}
\end{align}
where the $A_{X,\alpha}$ are the reduced matrix elements from which the WEFT coefficients have been factored out. In other words, $A_{X,\alpha}$ gives the amplitude for a neutrino to produce a charged lepton of flavor $\alpha$ via an interaction of type $X$, assuming this interaction is as strong as SM weak interactions. We discuss these amplitudes in detail below.  Note that \emph{no} summation over $\alpha$ is implied in \cref{eq:Mdecomposition}. For anti-neutrinos \cref{eq:Mdecomposition} holds after changing $\epsilon \to \epsilon^*$, and replacing $A_{X,\alpha}$ by the corresponding anti-neutrino amplitudes.  The total neutrino--nucleon scattering cross section is
\begin{align}
    \sigma_{\alpha\beta}^\text{tot} 
    &= \frac{1}{2 \kappa} \sum_\text{spins} \int\!d\Pi \, |\cM_{\alpha \beta}|^2 \nonumber\\
    &= \hat\sigma_{LL,\alpha} \delta_{\alpha\beta}
          + \sum_{X} \, [\epsilon_X ]_{\alpha\beta} \hat\sigma^\text{int}_{LX,\alpha} \delta_{\alpha\beta}
          + \sum_{X} \, [\epsilon_X ]^{*}_{\alpha\beta} (\hat\sigma^\text{int}_{LX,\alpha})^* \delta_{\alpha\beta}
          + \sum_{X,Y} \, [\epsilon_X ]_{\alpha\beta} [\epsilon_Y]^{*}_{\alpha \beta}
            \hat\sigma^\text{NP}_{XY,\alpha} \,,
    \label{eq:TotalXSection}
\end{align}
where the definition $\kappa = 4 E_\nu m_{n,p}$ has been introduced, with $m_{n,p}$ the mass of the target nucleon. Moreover, $\int\!d\Pi = \int\!\big[ \prod_j d^3p_j/(2\pi)^3 \big] \times (2\pi)^4 \delta^{(4)}(p_i - p_f)$ denotes the phase space integral. The integral runs over the three-momenta of all final-state particles (indexed by $j$), and the $\delta$-function ensures energy momentum conservation, with $p_i$ the total initial-state momentum and $p_f$ the total final-state momentum. Note that on the right-hand side of \cref{eq:TotalXSection}, again no sum over flavor indices is implied. Setting $\epsilon_X = 0$ recovers the SM cross section,
\begin{align}
    \hat\sigma_{LL,\alpha} \equiv \frac{1}{2 \kappa} \sum_\text{spins}
        \int\!d\Pi \, |A_{L,\alpha}|^2 \,. 
\end{align}
The terms describing the interference between the new physics and the SM amplitude are
\begin{align}
    \hat\sigma^\text{int}_{LX,\alpha} \equiv \frac{1}{2 \kappa} \sum_\text{spins}
        \int\!d\Pi (A_{L,\alpha} A^*_{X,\alpha}) \,, 
\end{align}
while the terms quadratic in the new physics interactions are
\begin{align}
    \hat\sigma^\text{NP}_{XY,\alpha} \equiv \frac{1}{2 \kappa} \sum_\text{spins}
        \int\!d\Pi (A_{X,\alpha} A^*_{Y,\alpha}) \,. 
\end{align}
In these relations the pre-factor of $\tfrac{1}{2}$ comes from averaging over initial state spins.. 

Let us now compute the reduced matrix elements for the CCQE interaction $\nu_\beta + n \to \ell_\alpha^- + p^+$. They can be expressed as 
\begin{align}
  \label{eq:PION_AP1}
    \begin{split}
        A_{L,\alpha}
            &= -\frac{2V_{ud}}{v^2} \Big[
                    \bar{u}_{\ell_\alpha}(p_{\ell_\alpha}) \gamma^\mu P_L u_\nu(p_\nu)
                    \Big]
                    \braket{p(p_p)|\bar{q}_u \gamma_\mu P_L q_d|n(p_n)} \,, \\
        A_{R,\alpha}
            &= -\frac{2V_{ud}}{v^2} \Big[
                    \bar{u}_{\ell_\alpha}(p_{\ell_\alpha}) \gamma^\mu P_L u_\nu(p_\nu)
                    \Big]
                    \braket{p(p_p)|{\bar q}_u \gamma_\mu P_R q_d|n(p_n)} \,, \\
        A_{S,\alpha}
            &= -\frac{V_{ud}}{v^2} \Big[
                    \bar{u}_{\ell_\alpha}(p_{\ell_\alpha}) P_L u_\nu(p_\nu) \Big]
                    \braket{p(p_p)|{\bar q}_u  q_d|n(p_n)} \,, \\ 
        A_{P,\alpha}
            &= \frac{V_{ud}}{v^2} \Big[
                   \bar{u}_{\ell_\alpha}(p_{\ell_\alpha}) P_L u_{\nu}(p_\nu) \Big]
                   \braket{p(p_p)|{\bar q}_u \gamma_5 q_d|n(p_n)}
        \,,  \\
        A_{T,\alpha}
            &= -\frac{V_{ud}}{2v^2} \Big[
                    \bar{u}_{\ell_\alpha}(p_{\ell_\alpha}) \sigma^{\mu\nu} P_L u_\nu(p_\nu)
                    \Big]
                    \braket{p(p_p)|{\bar q}_u \sigma_{\mu\nu} q_d|n(p_n)} \,,
    \end{split}
\end{align}
where the $u_i$ are the spinor wave functions of the leptons and hadrons (not to be confused with the up-quark field, $u$). The reduced matrix elements for the corresponding anti-neutrino interaction $\bar\nu_\beta + p \to \ell_\alpha^+ + n$, are obtained by replacing $u_{\ell_\alpha} \to v_\nu$, $u_\nu \to v_{\ell_\alpha}$, and taking the complex conjugate of the hadronic current as well as the CKM element $V_{ud}$. Note that, as long as neutrino masses are neglected, these amplitudes do no longer depend on the neutrino flavor as this dependence has been factored out in \cref{eq:Mdecomposition}. The hadronic matrix elements can be parameterized in terms of a set of Lorentz-invariant form factors, one corresponding to each possible quark bilinear~\cite{Weinberg:1958ut, Cirigliano:2013xha}:
\begin{align} 
    \braket{p(p_p)|{\bar q}_u \gamma_\mu q_d|n(p_n)}
        &= \bar u_p(p_p) \bigg[ G_V(Q^2) \gamma_\mu
                              + i \frac{\tilde{G}_{T(V)}(Q^2)}{2M_N} \sigma_{\mu\nu}q^\nu
                              - \frac{\tilde{G}_S(Q^2)}{2M_N}q_\mu
                         \bigg] u_n(p_n) \,,
    \label{eq:vectorC}\\
    \braket{p(p_p)|{\bar q}_u \gamma_\mu \gamma_5 q_d|n(p_n)}
        &= \bar u_p(p_p) \bigg[ G_A(Q^2) \gamma_\mu \gamma_5
                              + i \frac{\tilde{G}_{T(A)}(Q^2)}{2M_N} \sigma_{\mu\nu}q^\nu \gamma_5
                              - \frac{\tilde{G}_P(Q^2)}{2M_N} q_\mu \gamma_5
                         \bigg] u_n(p_n)\,,
    \label{eq:axialC}\\
    \braket{p(p_p)|{\bar q}_u  q_d|n(p_n)}
        &= G_S(Q^2) \, \bar{u}_p(p_p) u_n(p_n) \,,
    \label{eq:scalarC} \\[0.4cm]
    \braket{p(p_p)|{\bar q}_u \gamma_5 q_d|n(p_n)}
        &= G_P(Q^2) \, \bar u_p(p_p) \gamma_5 u_n(p_n) \,,
    \label{eq:pseudoscalarC} \\[0.2cm]
    \braket{p(p_p)|{\bar q}_u \sigma_{\mu\nu} q_d|n(p_n)}
        &= \bar u_p(p_p) \bigg[ G_T(Q^2) \sigma_{\mu\nu}
                              - \frac{i}{M_N} G_T^{(1)}(Q^2)
                                (q_\mu \gamma_\nu - q_\nu\gamma_\mu) \nonumber\\
        &\hspace{-1.5cm}      - \frac{i}{M_N^2} G_T^{(2)}(Q^2) (q_\mu P_\nu - q_\nu P_\mu)
                              - \frac{i}{M_N} G_T^{(3)}(Q^2)
                                (\gamma_\mu \dslash{q} \gamma_\nu
                               - \gamma_\nu \dslash{q} \gamma_\mu)
                         \bigg] u_n(p_n) \,,
    \label{eq:tensorC}
\end{align}
where $u_p$ and $u_n$ are the proton and neutron spinors, $q = p_p - p_n$ is the transferred momentum, $Q^2 \equiv -q^2$, $P = p_n + p_p$, and $M_N \equiv (m_n + m_p)/2$ is the average nucleon mass.

The crucial task now is to obtain the form factors $G_X$.
\begin{enumerate}
    \item {\bf Vector current:} In this case, there are three contributions, parameterized by the vector form factor $G_V(Q^2)$, the induced tensor-vector form factor $\tilde{G}_{T(V)}(Q^2)$, and the induced scalar form factor $\tilde G_S(Q^2)$, respectively.  In the literature, $G_V(Q^2)$ is sometimes called the Dirac form factor, and $\tilde{G}_{T(V)}(Q^2)$ is sometimes called the Pauli form factor, anomalous magnetic moment form factor, or weak magnetism form factor.  Naively, all terms in \cref{eq:vectorC} could be expected to be $\mathcal{O}(1)$. (The prefactors $\sim \mathcal{O}(q/M_N)$ accompanying $\tilde G_S(Q^2)$ and $\tilde{G}_{T(V)}(Q^2)$ are $\mathcal{O}(1)$ for GeV-scale neutrino scattering.) However, in the isospin-symmetric limit (neglecting the differences between the up and down quark masses), $\tilde G_S(Q^2)$ vanishes, as can be seen by replacing $p_p \leftrightarrow p_n$ and showing that this leads to a sign change in the induced scalar current. Isospin symmetry thus requires $\tilde G_S(Q^2) = 0$. (In the language of ref.~\cite{Weinberg:1958ut}, the induced scalar current is a second-class current which vanishes in the isospin-symmetric limit). Corrections to isospin symmetry are of $\mathcal{O}(10^{-3})$ and can be safely neglected here in view of uncertainties on other terms that are much larger.
    
    The vector [$G_V(Q^2)$] and induced tensor-vector [$\tilde{G}_{T(V)}(Q^2)$] form factors are related to the electromagnetic form factors of the nucleon $N=p,n$ by~\cite{Giunti:2007ry}
    \begin{align}
        G_V(Q^2)             &= F_1^p(Q^2) - F_1^n(Q^2) \,,
                                        \label{eq:VFF1} \\
        \tilde G_{T(V)}(Q^2) &= F_2^p(Q^2) - F_2^n(Q^2) \,.
                                        \label{eq:VFF2}
    \end{align}
    $F_1^N(Q^2)]$ and $F_2^N(Q^2)$, in turn can be related to the Sachs electric and magnetic form factors: 
    \begin{align}
        \begin{split}
            G_E^p(Q^2) &= F_1^p(Q^2)-\frac{Q^2}{4M_N^2}F_2^p=G_D(Q^2)\,,\\
            G_E^n(Q^2) &= F_1^n(Q^2)-\frac{Q^2}{4M_N^2}F_2^n=0\,,\\[0.2cm]
            G_M^p(Q^2) &= F_1^p(Q^2)+F_2^p(Q^2)=\mu_p G_D(Q^2)\,, \\[0.5cm]
            G_M^n(Q^2) &= F_1^n(Q^2)+F_2^n(Q^2)=\mu_n G_D(Q^2)\,,
        \end{split}
        \label{eq:Sachs}
    \end{align}
    where $\mu_p = 2.79284$ is the proton magnetic moment in units of the nuclear magneton, and $\mu_n = -1.91304$ is the equivalent for neutrons~\cite{ParticleDataGroup:2016lqr}. Finally, we can parameterize $G_D(Q^2)$ as a dipole function,
    \begin{align}
        G_D(Q^2) = \frac{1}{\big( 1 + \frac{Q^2}{M_V^2} \big)^{-2}} \,,
    \end{align}
    with $M_V = \SI{0.84}{GeV}$.

    Precise knowledge of the nucleon vector form factors can be obtained through high-statistics electron scattering experiments.
    \item {\bf Axial current:} Of the three terms contributing to the axial current, one -- namely the induced tensor-axial form factor, $\tilde{G}_{T(A)}$ -- vanishes in the isospin-symmetric limit \cite{Weinberg:1958ut}. The nucleon axial form factor, $G_A$, which is not accessible in electron-scattering experiments, introduces significant systematic uncertainties in neutrino--nucleus cross-section calculations. Existing experimental constraints on $G_A$ primarily rely on beta decay measurements, on neutrino scattering on nuclear targets, and on pion electro-production. However, the accuracy of these measurements remains relatively poor compared to measurements of the vector form factors. Beta decay experiments are sensitive only to the absolute normalization of the nucleon axial coupling, $g_A \equiv G_A(0)$, while neutrino scattering and pion electro-production experiments face challenges due to limited statistics and uncertainties in the nuclear modeling. 

    Historically, the most common parametrization used for the axial form factor is the dipole form:
    \begin{align}
        G_A(Q^2)=\frac{g_A}{\big(1 + \frac{Q^2}{m_A^2} \big)^2} \,,
        \label{eq:axialDip}
    \end{align}
    where the axial mass is $m_A \simeq 0.961$ GeV, used by GENIEv3 neutrino generator for their nominal $10a_02_11a$ model ~\cite{Andreopoulos:2009rq}) and $g_A = 1.2728 \pm 0.0017$~\cite{Gonzalez-Alonso:2018omy}.  This parametrization, however, proves incapable of capturing the observed shape of the axial form factor. The so-called $z$-expansion parametrization \cite{Simons:2022ltq},
    \begin{align}
        G_A(Q^2) = \sum_{k=0}^{\infty} a_k \, [z(Q^2)]^k\,,
        \label{eq:z-expansion}
    \end{align}
    substantially improves the accuracy of the fit to data \cite{Bhattacharya:2011ah, Meyer:2016oeg}. Here, $z(q^2)$ is an analytic function \cite{Hill:2006ub, Hill:2010yb, Meyer:2016oeg} motivated by QCD assumptions, while the coefficients $a_k$ can be obtained from lattice QCD or by fitting to neutrino--nucleus scattering and/or pion electro-production data.

    The different parametrizations of the axial form factor -- the dipole (dip) form factor from \cref{eq:axialDip}, the $z$-expansion fit to neutrino--deuteron scattering data (D2), and the $z$-expansion fit to lattice QCD results -- differ substantially, which reflects the large uncertainty the axial form factor introduces to the cross section. The top panel of \cref{fig:FormFactors} compares the different parametrizations and reveals discrepancies by almost an order of magnitude at large momentum transfer. This illustrates how important it is to quantitatively assess these systematic uncertainties. For the SM case such a study has been carried out in Ref.~\cite{Simons:2022ltq}. In the present paper, we analyze for the first time how form factor uncertainties affect the study of new interactions beyond the SM.

    \begin{figure*}
        \centering
        \includegraphics[width=0.9\textwidth]{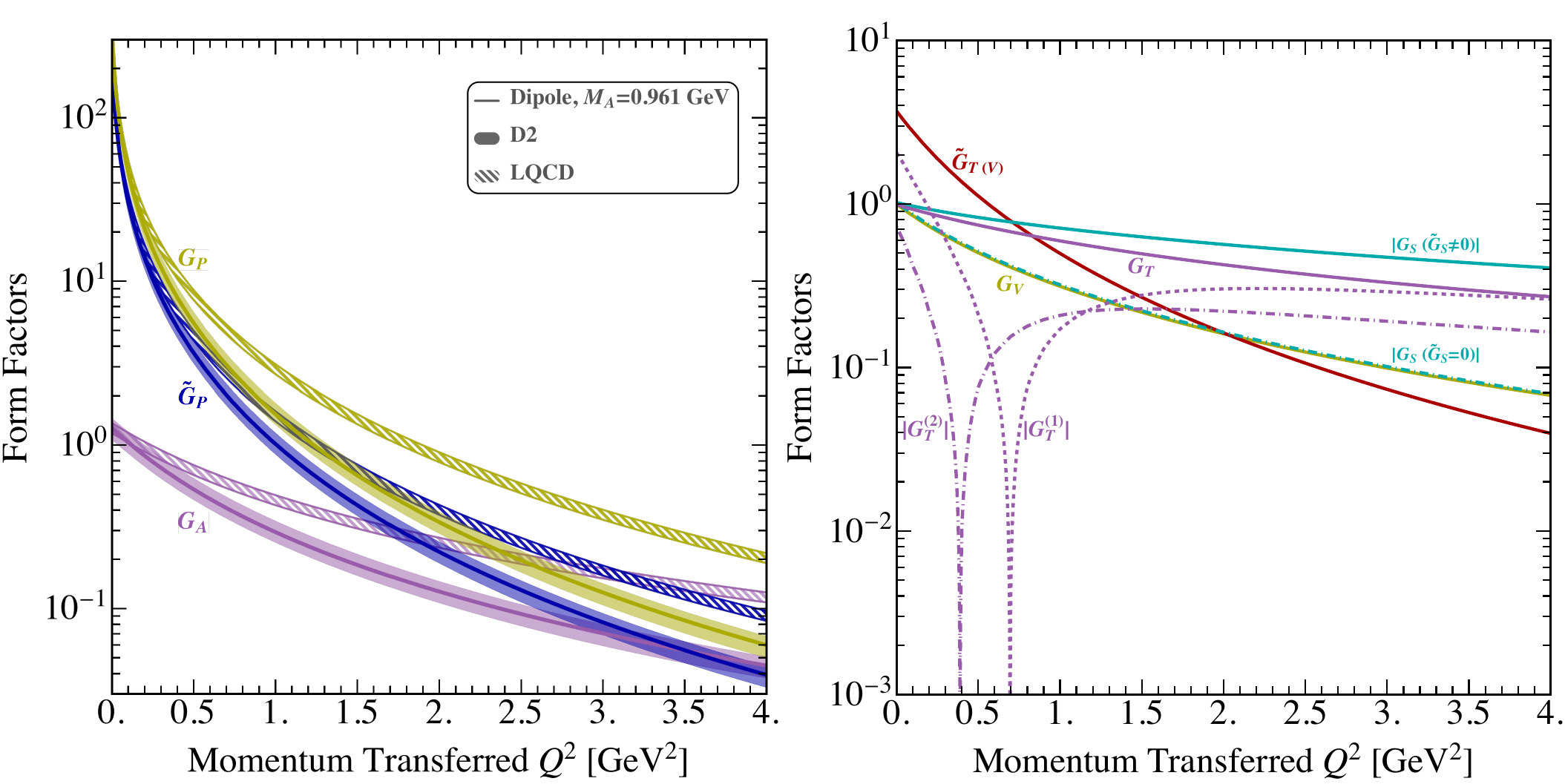}
        \caption{The nucleon form factors appearing in \cref{eq:vectorC,eq:axialC,eq:scalarC,eq:pseudoscalarC,eq:tensorC} as a function of $Q^2 = -q^2$.  For the form factors affected by the uncertainty in $G_A(Q^2)$, the left panel shows results for three different parametrizations, namely the dipole form factor with $m_A = \SI{0.961}{GeV}$ (thin solid lines), the $z$-expansion fitted to neutrino--deuterium scattering data (``D2'', shaded bands)~\cite{Meyer:2016oeg}, and a $z$-expansion fit to lattice QCD calculations by the RQCD Collaboration (hatched bands) \cite{RQCD:2019jai}. In the latter two cases, the width of the colored bands indicates the uncertainties quoted in the respective references. The remaining form factors are shown in the right panel. Note the sign change from negative to positive for $G_T^{(1)}$ and from positive to negative for $G_T^{(2)}$.}
        \label{fig:FormFactors}
    \end{figure*}

    \item {\bf Pseudoscalar current:} By using the partial conservation of the axial current (PCAC) and applying it to the axial current matrix element of the nucleon, one can find a relation between the axial, induced-pseudoscalar, and pseudoscalar form factors~\cite{Gonzalez-Alonso:2013ura}:
    \begin{align}
        G_P(Q^2) = \frac{M_N}{m_q} G_A(Q^2) + \frac{Q^2/2M_N}{2m_q} \tilde{G}_P(Q^2)\,.
        \label{eq:GP-pcac}
    \end{align}
    Here, $m_q = \SI{3.410(43)}{MeV}$ is the average light quark mass, taken from lattice calculations (2019 FLAG review, $N_f=2+1+1$ \cite{FlavourLatticeAveragingGroup:2019iem, EuropeanTwistedMass:2014osg, FermilabLattice:2018est}). Note the large prefactors proportional to $M_N/m_q$ and $Q^2 / (M_N m_q)$, which are both of order $10^3$. These prefactors lead to an enhancement of, e.g., $G_P(Q^2)$ compared to other form factors.
    
    We can moreover use the pion pole dominance (PPD) ansatz to derive an approximate relation between the induced-pseudoscalar and axial form factors. More precisely, we use the fact that $G_P(Q^2)$ is dominated by the pion-pole contribution because the pseudoscalar current has the same quantum numbers as the pion. This suggests to write $G_P(Q^2) = G_P(0) \, m_\pi^2 / (Q^2 + m_\pi^2)$, where $m_\pi$ is the pion mass. Moreover, in the low-$Q^2$ limit, \cref{eq:GP-pcac} implies $G_P(Q^2) \simeq (M_N / m_q) G_A(Q^2)$. Plugging both relations into \cref{eq:GP-pcac} leads to~\cite{Sasaki:2007gw, Falkowski:2021vdg, Meyer:2022mix}:
    \begin{align}
        \tilde G_P(Q^2) = -\frac{4M_N^2}{Q^2 + m_\pi^2} G_A(Q^2) \,.
        &\qquad\qquad\text{(low-$Q^2$ limit)}
    \end{align}
    For higher accuracy, we will instead use the $z$-expansion parametrization \cite{Simons:2022ltq} also for the pseudoscalar form factors, and write:
    \begin{align}
        \tilde G_P(Q^2) = -\frac{4M_N^2}{Q^2 + m_\pi^2}
                          \sum_{k=0}^{\infty} a^{\tilde{P}}_k \, z(Q^2)^k \,,
        \label{eq:z-expansion-GP} \,
    \end{align}
    and
    \begin{align}
        G_P(Q^2) = \frac{M_N}{m_q} \frac{m_\pi^2}{Q^2 + m_\pi^2}
                   \sum_{k=0}^{\infty} a^{P}_k \, z(Q^2)^k \,,
        \label{eq:z-expansion-GPtilde} \,
    \end{align}
    where the $a^{\tilde{P}}_k$ and $a^P_k$ coefficients are found by fitting to lattice QCD results. The first few coefficients in the expansion, taken from ref.~\cite{RQCD:2019jai}, are shown in \cref{tab:LQCDcoef}. We have checked that these lattice QCD results satisfy the PPD and PCAC relations.

\begin{table}
    \centering
    \begin{tabular}{c|ccccccc}
        \toprule
            \cmidrule{2-4} \cmidrule{5-7}
          $X$  & $a_0^X$      & $a_1^X$     & $a_2^X$ 
            & $a_3^X$ & $a_4^X$  & $a_5^X$  &$a_6^X$ \\
        \midrule
         $A$   & $1.009$ & $-1.756$    & $-1.059$  & $1.621$ & $3.919$     & $-5.739$ & $2.005$ \\
         $\Tilde{P}$   & $1.008$ & $-1.831$   &$-1.713$  & $4.994$ &  $-1.522$   & $-1.984$ & $1.047$ \\
         $P$ & $1.066$ & $-1.461$  & $-1.053$ & $-2.504$ & $12.446$   & $-12.260$ & $3.766$ \\
        \bottomrule
    \end{tabular}
    \caption{The coefficients of the $z$-expansion parametrization of the axial (first line) induced pseudoscalar (second line) and pseudoscalar (last line) form factors, determined by the lattice QCD calculations from Ref.~\cite{RQCD:2019jai}. }
    \label{tab:LQCDcoef}
\end{table} 

    \item {\bf Scalar current:} We can find the scalar form factor by using the conservation of the vector current (CVC), which relates the divergence of the vector current in (\cref{eq:vectorC}) to the scalar current (\cref{eq:scalarC}) and leads to \cite{Gonzalez-Alonso:2013ura}
    \begin{align}
        G_S(Q^2) = -\frac{\delta M_N^{QCD}}{\delta m_q} G_V(Q^2)
                   + \frac{Q^2/2M_N}{\delta m_q} \tilde G_S(Q^2) \,,
        \label{eq:GS}
    \end{align}
    where $\delta M_N^{QCD} = m_n - m_p = \SI{2.58(18)}{MeV}$ is the difference between the neutron and proton mass in pure QCD~\cite{Gonzalez-Alonso:2013ura}) and $\delta m_q=m_d-m_u=\SI{2.527(47)}{MeV}$ \cite{FlavourLatticeAveragingGroup:2019iem}.  Note that the induced tensor-vector form factor, $\tilde{G}_{T(V)}$, does not appear in \cref{eq:GS} because the divergence of the corresponding term in \cref{eq:vectorC} vanishes.
    
    The induced scalar form factor $\tilde G_S(Q^2)$ poses a problem as no robust measurements or lattice calculations for it exist. This is not surprising, given that, using Weinberg's language from ref.~\cite{Weinberg:1958ut}, it corresponds to a second-class current (that is, a current that violates $G$ parity). Such currents vanish in the isospin-symmetric limit, so their physical effects are always suppressed by small isospin-breaking factor such as $\delta M_N$ or $\delta m_q$. In \cref{eq:GS}, however, $\tilde G_S(Q^2)$ appears with a prefactor that is \emph{enhanced} by $1/\delta m_q$, compensating for the smallness of the form factor.
    
    In the absence of robust measurements or lattice calculations for $\tilde G_S(Q^2)$, we consider two approaches. The first is to simply set $\tilde G_S(Q^2) = 0$. The second one is based on the constituent quark model (CQM) which relates $\tilde{G}_S$ to the vector form factor, but suffers from large theoretical uncertainties. The CQM prediction is \cite{Sharma:2009hg}
    \begin{align}
        \tilde G_S(Q^2) = \frac{2 M_N}{(m_u + m_d)^{\text{CQM}}} \bigg(
                              \frac{\delta M_N^{QCD}}{2M_N} \, g_A
                            - \frac{\delta m_q}{(m_u+m_d)^{\text{CQM}}} \bigg) \, G_V(Q^2) \,,
    \end{align}
    where in the constituent quark model one takes $m_u = m_d = m_p/3$. 

    \item {\bf Tensor current:}
    Unlike the vector, axial-vector, scalar, and pseudoscalar currents, constraining the matrix elements of the antisymmetric tensor current is much more challenging.  While the relevant form factors are not easily determined through experimental data, Ward identities, or low-energy theorems, lattice QCD techniques and theoretical considerations can nevertheless provide valuable insights. In the following, we draw from Ref.~\cite{Hoferichter:2018zwu}. In the isospin-symmetric limit, which we adopt here, the $G^{(3)}_T(Q^2)$ form factor vanishes, and the following relation between the charged and neutral currents holds~\cite{FlavourLatticeAveragingGroup:2019iem}:
    \begin{align}
        \langle p (p_p)| \bar{u}\sigma^{\mu\nu}d|n(p_n)\rangle&=  \langle p (p_p)| \bar{u}\sigma^{\mu\nu}u- \bar{d}\sigma^{\mu\nu}d|p(p_p)\rangle\nonumber\\
        &=  \langle n (p_n)| \bar{d}\sigma^{\mu\nu}d- \bar{u}\sigma^{\mu\nu}u|n(p_n)\rangle\,.
    \end{align}
    The two equivalent matrix elements on the right-hand side of this equation can be evaluated on the lattice. The results can be written as
    \begin{align}
        G_T(Q^2)       &= F^u_{1,T}(Q^2) - F^d_{1,T}(Q^2) \,,\\[0.1cm]
        G_T^{(1)}(Q^2) &= F^u_{2,T}(Q^2) - F^d_{2,T}(Q^2) \,,\\[0.1cm]
        G_T^{(2)}(Q^2) &= F^u_{3,T}(Q^2) - F^d_{3,T}(Q^2) \,,
    \end{align}
    with the $F_{i,T}^q(Q^2)$ ($q=u,d$) functions parameterized by:
    \begin{align}
        F_{1,T}^q(Q^2) &= \pm \frac{1}{2} \Big[ F_{1,T}^u(0) - F_{1,T}^d(0)\Big]
                                              D_{b_1}(Q^2)
                            + \frac{1}{2} \Big[ F_{1,T}^u(0) + F_{1,T}^d(0)\Big]
                                              D_{h_1}(Q^2) \,,
                                                 \label{eq:F1T} \\[0.1cm]
        F_{2,T}^u(Q^2) &= -\frac{M_N}{2m_\pi} B_T^{\pi,u}(0)
                           \Big[ 2 G^p_M(Q^2) + G^n_M(Q^2) \Big]
                         + 2 \frac{M_N^2}{m^2_{b_1}} F_{1,T}^u(0) D_{b_1}(Q^2) \,,
                                                 \label{eq:F2Tu} \\
        F_{2,T}^d(Q^2) &= -\frac{M_N}{2m_\pi} B_T^{\pi,u}(0)
                           \Big[ G^p_M(Q^2) + 2 G^n_M(Q^2) \Big]
                         + 2 \frac{M_N^2}{m^2_{b_1}} F_{1,T}^d(0) D_{b_1}(Q^2) \,,
                                                 \label{eq:F2Td} \\
        F_{3,T}^u(Q^2) &=  \frac{M_N}{4m_\pi} B_T^{\pi,u}(0)
                           \Big[ 2 F^p_2(Q^2) + F^n_2(Q^2) \Big]
                         - \frac{M_N^2}{m^2_{b_1}} F_{1,T}^u(0) D_{b_1}(Q^2) \,,
                                                 \label{eq:F3Tu} \\
        F_{3,T}^d(Q^2) &=  \frac{M_N}{4m_\pi} B_T^{\pi,u}(0)
                           \Big[ F^p_2(Q^2) + 2 F^n_2(Q^2) \Big]
                         - \frac{M_N^2}{m^2_{b_1}} F_{1,T}^d(0) D_{b_1}(Q^2) \,.
                                                 \label{eq:F3Td}
    \end{align}
    In the first of these relations, the plus sign corresponds to $q=u$ and the minus sign to $q=d$. Numerically, Ref.~\cite{Hoferichter:2018zwu} gives $F_{1,T}^u(0) = 0.784$, $F_{1,T}^d(0) = -0.204$, and $B_T^{\pi,u}(0) = 0.195$ for the tensor charges that determine the normalization. The functions $G^N_M(Q^2)$ and $F_2^N(Q^2)$ are the Sachs magnetic and Pauli form factors defined in \cref{eq:Sachs}.  Finally,
    \begin{align}
        D_{b_1/h_1}(Q^2) = \frac{m^2_{b_1/h_1}}{m^2_{b_1/h_1} + Q^2}
        \label{eq:D-b1-h1}
    \end{align}
    are dipole functions with $m_{h_1} = \SI{1.166}{GeV}$ and $m_{b_1} = \SI{1.229}{GeV}$.
    
    Note that several of the terms in \cref{eq:F1T,eq:F2Tu,eq:F2Td,eq:F3Tu,eq:F3Td} benefit from an enhancement factor $M_N / m_\pi \sim 10$, which facilitates detection of tensor currents.
\end{enumerate}
The above parametrizations of the form factors -- especially the axial vector and tensor ones -- have been derived with momentum transfers $\sqrt{Q^2} \lesssim \SI{1}{GeV}$ in mind. However, given their weak dependence on $Q^2$ at $\mathcal{O}(\si{GeV})$ momentum transfers, which is evident from \cref{fig:FormFactors}, we will use them also to extrapolate to larger momentum transfers relevant for accelerator neutrino experiments.

We are now ready to express the spin-summed squared amplitudes for CCQE neutrino--nucleon scattering ($\nu_\beta + n \to \ell_\alpha^- + p^+)$ in terms of the form factors and as functions of the Mandelstam variables: 
\begin{align}
    \frac{1}{2} \sum_\text{spin} A_{X,\alpha} A^*_{Y,\alpha}
        = C_0 \bigg[ B_{\alpha,XY}(Q^2)
                   + C_{\alpha,XY}(Q^2) \frac{s-u}{M_N^2}
                   + D_{\alpha,XY}(Q^2) \frac{(s-u)^2}{M_N^4} \bigg] \,,
    \label{eq:AmpSq}
\end{align}
where $X,Y={L,R,S,P,T}$ and $C_0 \equiv 4 |V_{ud}|^2 M_N^4 / v^4$. The factor $\frac{1}{2}$ on the left hand side averages over spin orientations of the initial nucleon. In the approximation that the nucleon is initially at rest, kinematics dictates that $s - u = 4 M_N E_\nu - Q^2 - m_{\ell_\alpha}^2$.  Assuming moreover that proton and neutron masses are identical, the functions $B_{\alpha,XY}(Q^2)$, $C_{\alpha,XY}(Q^2)$ and $D_{\alpha,XY}(Q^2)$ are\footnote{The full expressions without these approximation are given in \cref{app:AmpSq}. For our numerical results, we will use the full expressions because nucleons bound in nuclei cannot be assumed to be initially at rest.}
\begin{align}
\cline{1-3}
%
    B_{\alpha,LL} &=
        \frac{m_{\ell_\alpha}^2 + Q^2}{M_N^2} \bigg\{
            \Big(1 + \frac{Q^2}{4M_N^2} \Big) G_A^2
          - \Big(1 - \frac{Q^2}{4M_N^2} \Big)
                \Big(F_1^2 -\frac{Q^2}{4M_N^2} F_2^2 \Big)
          + \frac{Q^2}{M_N^2} F_1 F_2 \nonumber\\
        &\quad
          - \frac{m_{\ell_\alpha}^2}{4M_N^2} \bigg[
                \Big( F_1 + F_2 \Big)^2
              + \Big( G_A - \tilde{G}_P \Big)^2
              - \Big( 1 + \frac{Q^2}{4M_N^2} \Big) \tilde{G}_P^2 \bigg]
            \bigg\}\,, \label{eq:B-LL} \\
    C_{\alpha,LL} &=
        \frac{Q^2}{M_N^2} \Big( G_A(F_1+F_2) \Big) \,, \\
    D_{\alpha,LL} &=
        \frac{1}{4} \Big(G_A^2 + F_1^2
                       + \frac{Q^2}{4M_N^2} F_2^2 \Big) \,, \\
\cline{1-3}
%
    B_{\alpha,RR} &=  B_{\alpha,LL} \,, \\[0.2cm]
    C_{\alpha,RR} &= -C_{\alpha,LL} \,, \\[0.2cm]
    D_{\alpha,RR} &=  D_{\alpha,LL} \,, \\[-0.1cm]
\cline{1-3}
%
    B_{\alpha,SS} &= \frac{m_{\ell_\alpha}^2 + Q^2}{M_N^2}
                     \Big(1 + \frac{Q^2}{4M_N^2} \Big) G_S^2 \,,\\
    C_{\alpha,SS} &= 0 \,, \\[0.2cm]
    D_{\alpha,SS} &= 0 \,, \\[-0.1cm]
\cline{1-3}
%
    B_{\alpha,PP} &= \frac{m_{\ell_\alpha}^2 + Q^2}{M_N^2}
                     \frac{Q^2}{4M_N^4} G_P^2\,, \\
    C_{\alpha,PP} &= 0 \,, \\[0.2cm]
    D_{\alpha,PP} &= 0 \,, \\[-0.1cm]
\cline{1-3}
%
    B_{\alpha,TT} &=
        - \frac{m_{\ell_\alpha}^2 + Q^2}{M_N^2} \bigg\{
            G_T^2 + \frac{Q^2}{4 M_N^2} \bigg[
                + 4 G_T G_T^{(1)}
                - \frac{m_{\ell_\alpha}^2}{M_N^2} \Big(
                    2 G_T^{(2)} \big(G_T - 4 G_T^{(2)} \big) 
                                                       \nonumber\\
        &\hspace{8cm}
                  + \big( G_T^{(1)} \big)^2
                  - 4 G_T^{(1)} G_T^{(2)} \Big) \bigg] \nonumber\\
        &\quad
         + \frac{m_{\ell_\alpha}^2}{2M_N^2} \bigg[
               G_T^2 - 4 G_T \Big(({G_T^{(1)}} + {G_T^{(2)}} \Big)
             + 2 \Big( G_T^{(1)} + 2 G_T^{(2)} \Big)^2 \bigg]
                               \nonumber\\
        &\quad
         + 4 \bigg( \frac{Q^2}{4M_N^2} \bigg)^2
             \Big( \big(G_T^{(1)}\big)^2
                 + \frac{m_{\ell_\alpha}^2}{M_N^2}
                   \big(G_T^{(2)}\big)^2 \Big)
        \bigg\}\,, \\
    C_{\alpha,TT} &= 0 \,, \\
    D_{\alpha,TT} &=
        \frac{1}{2} G_T^2 + \frac{Q^2}{4M_N^2} \bigg[
            -2 G_T G_T^{(2)} + \Big( G_T^{(1)} + 2 G_T^{(2)} \Big)^2
        \bigg]
      + 4{G_T^{(2)}}^2 \Big(\frac{Q^2}{4M_N^2}\Big)^2 \,, \\
\cline{1-3} \notag \\[-0.5cm]
\intertext{for $XX$-type interactions, and}
\cline{1-3} \notag
%
    B_{\alpha,LR} &=
        \frac{m_{\ell_\alpha}^2 + Q^2}{M_N^2} \bigg\{
            - \Big( 1 + \frac{Q^2}{4M_N^2} \Big) G_A^2
            - \Big( 1 - \frac{Q^2}{4M_N^2} \Big)
              \Big( F_1^2 -\frac{Q^2}{4M_N^2} F_2^2 \Big)
            + \frac{Q^2}{M_N^2} F_1 F_2 \nonumber\\
        &\quad
            -\frac{m_{\ell_\alpha}^2}{4M_N^2} \bigg[
                \Big( F_1 + F_2 \Big)^2
              - \Big( G_A - \tilde{G}_P \Big)^2
              + \Big( 1 + \frac{Q^2}{4M_N^2} \Big) \tilde{G}_P^2 \bigg]
            \bigg\} \,, \label{eq:B-LR} \\  
    C_{\alpha,LR} &= 0 \,, \label{eq:C-LR} \\
    D_{\alpha,LR} &= \frac{1}{4} \Big(-G_A^2 + F_1^2
                       + \frac{Q^2}{4M_N^2} F_2^2 \Big) \,, 
                       \label{eq:D-LR} \\
\cline{1-3}
%
    B_{\alpha,LS} &= 0 \,, \\
    C_{\alpha,LS} &= \frac{m_{\ell_\alpha}}{2M_N}
                     \Big( F_1 - \frac{Q^2}{4M_N^2}F_2 \Big) G_S \,, \\
    D_{\alpha,LS} &= 0 \,, \\[-0.1cm]
\cline{1-3}
%
    B_{\alpha,LP} &=
        -\frac{m_{\ell_\alpha}}{2M_N}
         \frac{m_{\ell_\alpha}^2+Q^2}{M_N^2}
         \Big( G_A + \frac{Q^2}{4M_N^2} G_P \Big) G_P \,,
                    \label{eq:B-LP} \\
    C_{\alpha,LP} &= 0 \,, \\[0.2cm]
    D_{\alpha,LP} &= 0 \,, \\[-0.1cm]
\cline{1-3}
%
    B_{\alpha,LT} &=
       - \frac{m_{\ell_\alpha}^2 + Q^2}{M_N^2}
        \frac{m_{\ell_\alpha}}{M_N} \bigg\{
             F_1 \Big( 3 G_T - 2 G_T^{(1)} - 4 G_T^{(2)} \Big)
            + 2 F_2 G_T
                            \nonumber\\
    &\quad
            + \frac{Q^2}{4M_N^2} \Big[
                  4 F_1 \big( G_T^{(1)} - G_T^{(2)} \big)
                + F_2 \big(-G_T + 6 G_T^{(1)} + 4 G_T^{(2)} \big) \Big]
            + 4 F_2 G_T^{(2)} \Big( \frac{Q^2}{4M_N^2} \Big)^2
        \bigg\} \,, \label{eq:B-LT} \\
    C_{\alpha,LT} &=
        -\frac{m_{\ell_\alpha}}{M_N} \Big( 3 G_A G_T
            + (\tilde{G}_P G_T - 4 G_A G_T^{(1)}) \frac{Q^2}{4M_N^2} \Big) \,,
                    \label{eq:C-LT} \\
    D_{\alpha,LT} &= 0 \,. \label{eq:D-LT}  \\
\cline{1-3} \notag
\end{align}
for the interference terms with the SM amplitude ($LX$-type interactions).

For anti-neutrinos interacting via $\bar\nu_\beta + p^+ \to \ell_\alpha^+ + n$,  the relevant functions for the LL, RR, LS and LT interactions are $({\bar{C}}_{\alpha,XY}) =-(C_{\alpha,XY})$, for the LP interaction we have $\bar{B}_{\alpha,LP} =-B_{\alpha,LP}$, and the remaining functions are identical to their $\nu$ counterparts. Finally, for both neutrino and anti-neutrino scatterings we always have $\sum_\text{spin} A^*_{X,\alpha} A_{Y,\alpha} = \sum_\text{spin}{A}_{X,\alpha} A^*_{Y,\alpha}$.

Let us focus first on the interference ($LX$-type) terms which are typically more important than the $XX$-type contributions because they will be multiplied by only one power of the new physics couplings $\epsilon_X$. We see that among the interference terms, the only ones that are not proportional to $m_{\ell_\alpha} / M_N$ are of the $LR$ type (interference between the left-handed SM current and a right-handed new physics current, \cref{eq:B-LR,eq:C-LR,eq:D-LR}). Therefore, for electron (muon) neutrinos, for which $m_{\ell_\alpha} / M_N \sim \num{5e-4}~(0.1)$, one might expect right-handed new physics currents parameterized by $\epsilon_R$ to be the easiest to detect. However, it turns out that the $F_1$, $F_2$, and $G_S$ form factors appearing in \cref{eq:B-LR,eq:D-LR} are only $\mathcal{O}(1)$, while tensor (\cref{eq:B-LT,eq:C-LT}) and pseudoscalar interactions (\cref{eq:B-LP}) benefit from form factors that are larger (see discussion below \cref{eq:GP-pcac} and below \cref{eq:D-b1-h1}), compensating the smallness of  $m_{\ell_\alpha} / M_N$.

At quadratic order in the $\epsilon_X$'s we do not have any such suppressions and all interactions contribute, especially the pseudoscalar and tensor ones with their large form factors.

\subsection{Neutrino--Nucleus Scattering}
\label{sec:nu-nucleus}

While interactions of neutrinos with nucleons are already highly non-trivial, the fact that nucleons are embedded in nuclei adds another layer of complexity.  In this section we consider the scattering of (anti)neutrinos off a target nucleus $A$. We will focus in particular on $A=16$ (oxygen, relevant for water \v{C}erenkov detectors), for which nuclear spectral functions are available.\footnote{Our results should to a good approximation be applicable also to scattering on carbon ($A=12$) and nitrogen ($A=14$), given that these nuclei are similar in mass to oxygen.  We plan to repeat our analysis for scattering on ${}^{40}$Ar once the corresponding spectral functions -- currently under development -- are published.}

In the limit of moderate momentum transfer ($|\vec{q}| \gtrsim \SI{400}{MeV}$),
a factorization scheme can be adopted to describe the hadronic final state~\cite{Benhar:2006wy}. In the quasielastic region, the latter can be simplified as the product of a single nucleon with momentum $p'=(E',{\vec{p}'})$ that is decoupled, and the $A-1$-nucleon residual system:
\begin{align}
    \ket{\Psi_f^A} = \ket{p'} \otimes \ket{\Psi^{A-1}_{n},p_{A-1}} \,.
    \label{eq:fact:1b}
\end{align}
Here, $\ket{p'}$ is the final state nucleon produced at the primary vertex, assumed to be in a plane wave state, and $\ket{\Psi^{A-1}_{n},p_{A-1}}$ is the wave-function describing the remnant nucleus with momentum $p_{A-1}$. The subscript $n$ labels the possible states of the nuclear remnant, with $n=0$ corresponding to the ground state and $n>0$ to excited states. With this factorization ansatz, and inserting a complete set of free single-nucleon states satisfying $\sum_N \int d^3p_N \ket{N,p_N} \bra{p_N,N} = 1$ (where the sum runs over all nucleons in the target nucleus and the integral runs over the nucleon 3-momentum), the matrix element of the one-body current operator is
\begin{align}\label{eq:1bodyME}
    \bra{\Psi_f^A} J^\mu \ket{\Psi^A_0} = \sum_N \int d^3p_N
        \big[\!\bra{\Psi^{A-1}_{n}} \otimes \bra{p_N,N} \big] \ket{\Psi^A_0}
        \bra{p'}\sum_{i}j^\mu_{i} \ket{p_N,N} \,.
\end{align}
In this expression $\ket{\Psi^A_0}$ is the ground state of the initial-state nuclear system, $p'=q+p_N$ is the final 4-momentum of a free nucleon with initial 4-momentum $p_N$, and $J^\mu=\sum_{i} j^\mu_{i}$, is the nuclear current operator, which we write as a sum of one-body currents, each of them corresponding to one of the quark bilinear operators from \cref{eq:vectorC,eq:axialC,eq:scalarC,eq:pseudoscalarC,eq:tensorC}. 

Exploiting momentum conservation at the single nucleon vertex we can rewrite the squared spin-summed amplitudes in the nuclear case. Considering the system where a nucleon with momentum $p_N$ is removed, leaving behind a residual nucleus with excitation energy $E^*$,\footnote{In other words, $E^*$ is the absolute value of the energy required to remove one nucleon from the nucleus.} we have 
\begin{align}
    \frac{1}{2} \sum_\text{spin} \mathcal{A}_{X,\alpha} \mathcal{A}^*_{Y,\alpha}  &=
        \frac{1}{2} \int \frac{d^3 p_N}{(2\pi)^3} P_h({\vec{p}_N, E^*})
            \frac{m_N}{e(\vec{p}_N)} \frac{m_N}{e(\vec{q} + \vec{p}_N)} \nonumber\\[0.2cm]
    &\quad\times \sum_\text{spin} \sum_N
        A_{X,\alpha} A^*_{Y,\alpha} \,
        \delta\big( \tilde\omega + e(\vec{p}_N) - e(\vec{q} + {\vec{p}_N}) \big)\,,
        \label{eq:squaredM-nuclear}
\end{align}
where $e(\vec{p}_N) \equiv (\vec{p}_N^2 + m_N^2)^{1/2}$, and $\tilde\omega$ is the effective energy transfer from the neutrino to the nucleon. It is defined in terms of the physical energy transfer $\omega$ (the zero component of the 4-vector $q \equiv (\omega, \vec{q})$) via
\begin{align}
    \tilde\omega &= \omega + m_N - E^* - e(\vec{p}_N) \,.
    \label{eq:omega-tilde}
\end{align}
The correction terms on the right-hand-side of \cref{eq:omega-tilde} account for the fact that the initial state nucleon is bound. The one-nucleon spectral function, $P_h(\vec{p}_N, E)$, can be written as 
\begin{align}
    P_{h}(\vec{p}_N,E^*) = \sum_n \Big| \bra{\Psi^A_0} \big[ \ket{p_N} \otimes \ket{\Psi^{A-1}_n}\big] \Big|^2\,
                         \delta(E^* + E_0^A - E_n^{A-1}+m_N) \,,
\end{align}
where the energy of the nuclear ground state is $E_0^A$ and the one of the remnant system, which can be found in any state $n$, is $E_n^{A-1}$. The factors $m_{N}/e({\bf p_N})$ and $m_{N}/e(\textbf{q}+{\bf p_N})$ in \cref{eq:squaredM-nuclear} are included to account for the fact that the four-spinors in the relativistic matrix element are defined using the covariant normalization, whereas the spinors appearing in the spectral function are normalized to unity.

Various methods can be employed to calculate the spectral function of finite nuclei~\cite{Dickhoff:2004xx, Barbieri:2016uib, Barbieri:2019ual, Sobczyk:2023mey, CLAS:2022odn, Andreoli:2021cxo}. In this study, we utilize the spectral function of ${}^{16}$O from Ref.~\cite{Benhar:1994hw,Benhar:1989aw}. This approach employs a semi-phenomenological method, combining $(e,e'p)$ scattering data with nuclear matter calculations derived from the Correlated Basis Function formalism.

\section{Cross Section Results}
\label{sec:XSections}

Using the formalism outlined in the previous sections, we can now calculate the CCQE neutrino--nucleus cross-section, including the full set of EFT operators. In the laboratory frame, where the initial nucleus is at rest, the differential neutrino--nucleus cross-section is given by
\begin{align}
    \frac{d\sigma_{\alpha\beta}}{dQ^2}
         &= \frac{d\hat\sigma_{LL,\alpha}}{dQ^2} \delta_{\alpha\beta}
          + \sum_X \Big(
                [\epsilon_X]_{\alpha\beta} \frac{d\hat\sigma_{LX,\alpha}}{dQ^2} \delta_{\alpha\beta}
              + h.c. \Big)
          + \sum_{X,Y,\beta}
                [\epsilon_X]_{\alpha\beta} \, [\epsilon_Y]^*_{\alpha\beta}
                \frac{d\hat\sigma_{XY,\alpha}}{dQ^2} \,,
    \label{eq:DifferentialXSection}
\end{align}
with
\begin{align}
    \frac{d\hat\sigma_{XY,\alpha}}{dQ^2} &=
        \frac{1}{64\pi} \frac{1}{(p_\nu \cdot p_{N_i})^2} \bigg[
            \frac{1}{2} \sum_\text{spin} 
        \mathcal{A}_{X,\alpha} \mathcal{A}^*_{Y,\alpha} \bigg] \,.
\end{align}
Here, $N_i = n$ for neutrino interactions ($\nu n$ scattering) and $N_i=p$ for anti-neutrino interactions ($\bar\nu p$ scattering). For the scattering on free nucleons, one simply has to replace $\mathcal{A}_{X,\alpha} \mathcal{A}^*_{Y,\alpha}$ by $A_{X,\alpha} A^*_{Y,\alpha}$. From kinematics we have  $p_\nu \cdot p_{N_i} = M_N E_\nu$. The total cross section (\cref{eq:TotalXSection}) is obtained by integrating \cref{eq:DifferentialXSection} over $Q^2$, with the lower and upper integration limits \cite{Workman:2022ynf}
\begin{align}
    Q^2_\text{min} &= \frac{1}{2 E_\nu + M_N} \bigg[
        2 M_N E_\nu^2 - m_{\ell_\alpha}^2 (M_N + E_\nu)
      - E_\nu \sqrt{-2 M_N^2 (s + m_{\ell_\alpha}^2) + (s - m_{\ell_\alpha}^2)^2 + M_N^4}
    \bigg] \,, \\
    Q^2_\text{max} &= \frac{1}{2 E_\nu + M_N} \bigg[
        2 M_N E_\nu^2 - m_{\ell_\alpha}^2 (M_N + E_\nu)
      + E_\nu \sqrt{-2 M_N^2 (s + m_{\ell_\alpha}^2) + (s - m_{\ell_\alpha}^2)^2 + M_N^4}
    \bigg] \,.
\end{align}
The squared center of mass energy is $s = M_N^2 + 2 M_N E_\nu$.

\begin{figure*}
    \centering
    \includegraphics[width=0.85\textwidth]{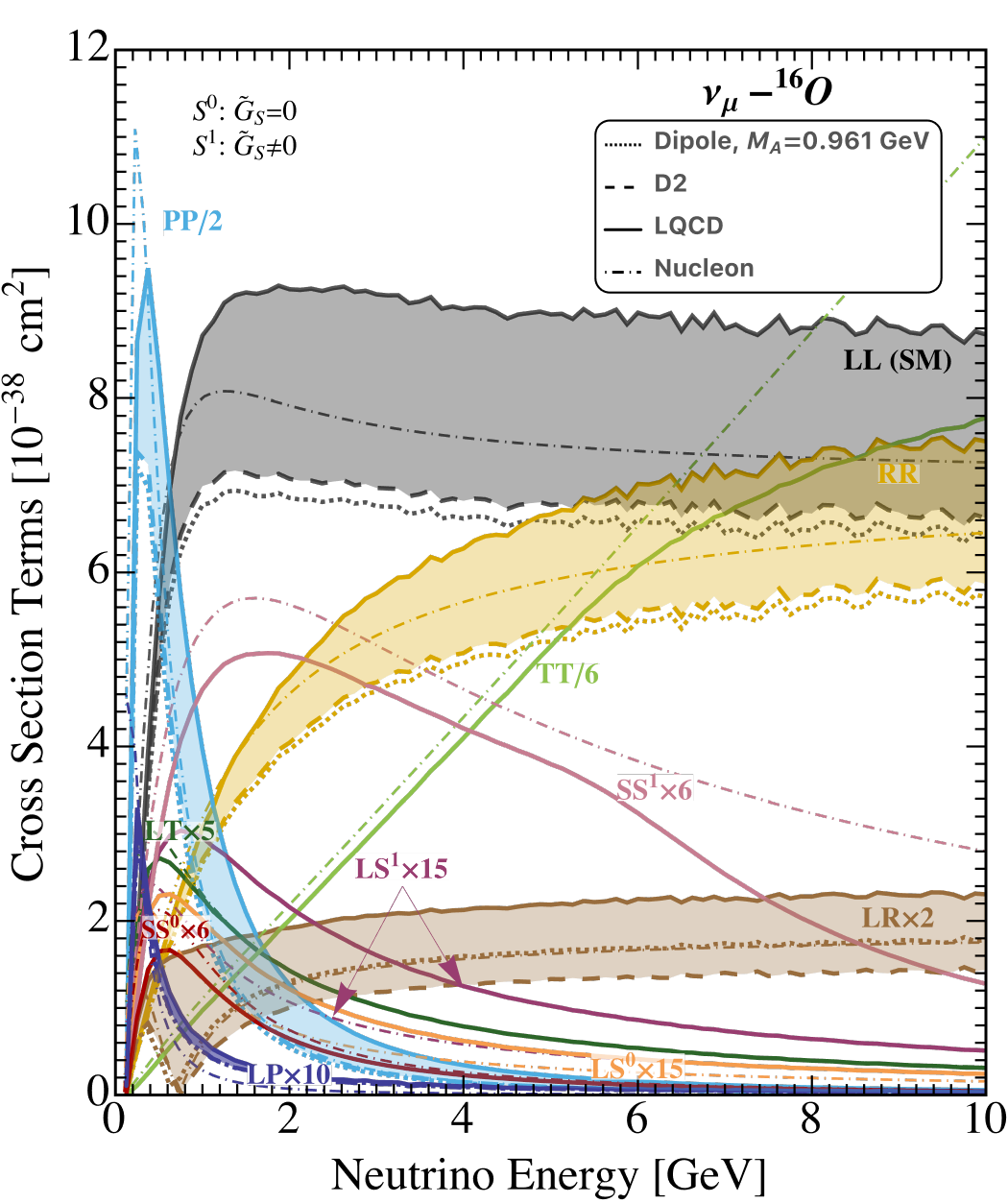}
    \caption{Contributions to the CCQE differential cross sections for muon neutrinos scattering on an oxygen target, as a function of the neutrino energy. Results for $\nu_e$ scattering are very similar. The different colored curves correspond to operators with different Lorentz structures, with the SM ($LL$) case shown in gray. For interactions depending on the axial form factor, we compare different parameterizations of that form factor: the dipole from \cref{eq:axialDip} (dotted), the $z$-expansion fitted to neutrino--deuteron scattering data (dashed), and the $z$-expansion fitted to lattice QCD results (solid). For comparisons, we also show results for neutrino scattering on free nucleons (thinner dot-dashed lines). The content of this plot is available in tabulated form from the Our main cross-section results in tabulated form are available from \href{https://github.com/ztabrizi/EFT-in-Neutrino-Nucleus-Scattering/}{GitHub} \cite{github}.}
    \label{fig:AllXSections-mu}
\end{figure*}

\begin{figure*}
    \centering
    \includegraphics[width=0.85\textwidth]{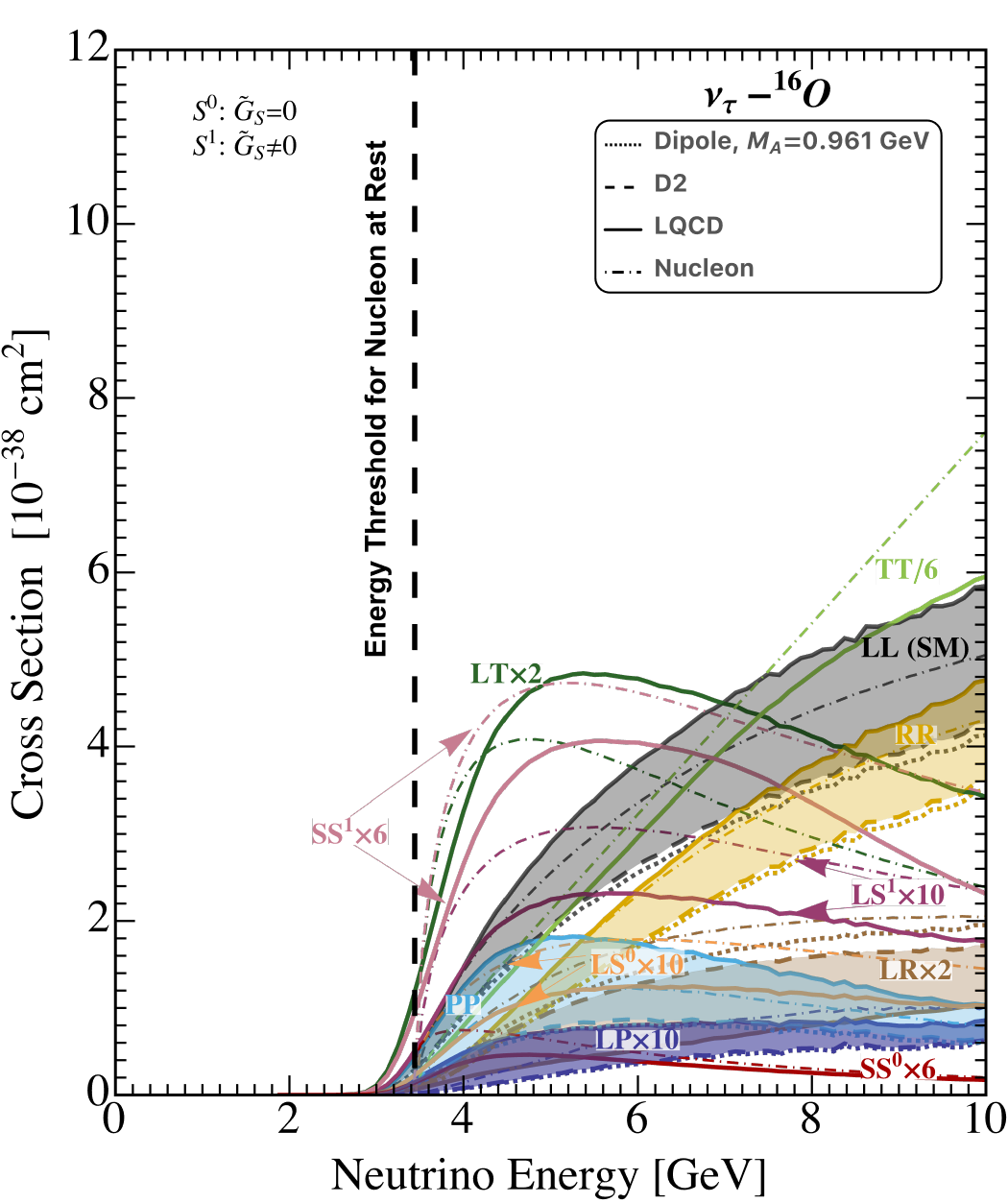}
    \caption{Same as \cref{fig:AllXSections-mu}, but for $\nu_\tau$ scattering.}
    \label{fig:AllXSections-tau}
\end{figure*}

The individual contribution, $\hat\sigma_{XY,\alpha}$, to the total cross sections as a function of the neutrino energy are shown in \cref{fig:AllXSections-mu,fig:AllXSections-tau} for scattering on oxygen.  Remember that the $\hat\sigma_{XY,\alpha}$ can be negative -- only after multiplying with the $[\epsilon_X]_{\alpha\beta}$ coefficients and summing according to \cref{eq:TotalXSection}, the result for the total cross section needs to be positive. In fact, $\sigma_{LR}$ exhibits a sign change (from negative at lower energies to positive at higher energies) at $E_\nu \sim \SI{0.8}{GeV}$. 

\sisetup{detect-weight=true,detect-inline-weight=math}
\begin{table}[t]
    \centering
    \scalebox{0.8}{
    \begin{tabular}{ll|ccc|ccc|cc}
        \toprule
         
        & &  \multicolumn{3}{c|}{$\nu_e$} &
        \multicolumn{3}{c|}{$\nu_\mu$}  & \multicolumn{2}{c}{$\nu_\tau$} \\
        \midrule
        \midrule
        & $E_\nu ~[{\rm{GeV}}]$ &  $\leq 1$ & $1<\cdots\leq4$ & $>4$& $\leq 1$ & $1<\cdots\leq4$& $>4$ & $\leq4$ & $>4$\\
        \midrule
       & LL       
        &2.4
        &3.1
        &3.0
        &2.3
        &3.1
        &3.0
        &3.4
        &3.4
        \\
       & RR    
        &1.4
        &2.1
        &2.6
        &1.4
        &2.1
        &2.6
        &0.2
        &2.8
        \\
      {D2 Difference ($\%$) }& PP     
        &4.9
        &7.5
        &7.8
        &4.5
        &7.2
        &7.5
        &12.1
        &6.1
        \\
        &LR    
        &10.4
        &23.5
        &19.6
        &9.1
        &23.5
        &19.6
        &8.0
        &11.7
        \\
       & LP   
        &1.5
        &0.04
        &0.2
        &1.4
        &0.9
        &0.05
        &9.4
        & 4.1\\
        \midrule
         & LL       
        &18.3
        &34.9
        &35.7
        &17.8
        &34.9
        &35.7
        &70.0
        &44.5
        \\
       & RR    
        &11.8
        &22.4
        &30.5
        &11.7
        &22.4
        &30.5
        &30.8
        &37.0
        \\
      {LQCD Difference ($\%$)}& PP     
        &60.5
        &145.0
        &162.7
        &56.5
        &138.6
        &154.6
        &310.6
        &90.4
        \\
        &LR    
        &76.8
        &167.5
        &176.3
        &76.3
        &168.1
        &176.5
        &154.1
        &152.3
        \\
       & LP   
        &14.0
        &10.5
        &4.1
        &16.3
        &17.3
        &10.9
        &974.8
        &86.0 \\
         \bottomrule
    \end{tabular}
    }
    \caption{Differences (in per cent) between the energy-integrated cross sections evaluated using axial form factors based on neutrino--deuteron scattering (top) or lattice QCD (bottom) compared to the simple dipole form factor from \cref{eq:axialDip}. Different columns correspond to different energy ranges and neutrino flavors, while different rows are for interactions with different Lorentz structures.}
    \label{tab:Uncertainties}
\end{table}

Our main findings from \cref{fig:AllXSections-mu,fig:AllXSections-tau} can be summarized as following:
\begin{enumerate}
    \item New interactions can benefit from significant cross-section enhancements, aiding their detection. This is particularly true for pseudoscalar and tensor interactions, which benefit from large nucleonic form factors.  In the case of pseudoscalar currents, this is related to the fact that the pion is a pseudoscalar, and nucleons couple strongly to pions. For the tensor case, the enhancement can similarly be related to the existence of two-pion intermediate states \cite{Hoferichter:2018zwu}.

    \item The axial form factor introduces significant systematic uncertainties to the neutrino--nucleus cross-section, indicated by the colored bands in \cref{fig:AllXSections-mu,fig:AllXSections-tau} and also summarized in \cref{tab:Uncertainties}. This affects the SM contribution as well as many types of new interactions, notably those with right-handed and pseudoscalar Lorentz structures.  We see that using lattice QCD, which predicts a larger form factor, as input leads to cross sections that are larger by $\mathcal{O}(10\%)$ factor than those based on the dipole form factor or neutrino--deuteron scattering. Discrepancies are largest in the multi-GeV energy range which is most relevant to long-baseline experiments like DUNE. Note, however, that the large uncertainties affect mostly the normalization of the cross section, not the energy dependence. Therefore, the energy spectrum of observed neutrino events can be used to distinguish different types of new interactions.

    \item Nuclear effects are crucial even at multi-GeV energies, in contradiction to the widespread assumption that neutrino scattering on free nucleons is a good approximation in the calculation of the CCQE cross sections at large $q^2$. This is particularly apparent for tensor interactions (green lines in \cref{fig:AllXSections-mu,fig:AllXSections-tau}) at energies $\gtrsim \SI{6}{GeV}$. The observed phenomenon can be attributed to the Fermi motion of nucleons within the nuclear medium. This introduces additional terms that are negligible when scattering occurs on a single nucleon at rest in the laboratory frame.
    
    \item For $\tau$ neutrinos, the initial-state momentum of the bound target nucleon lowers the energy threshold for $\nu_\tau$ interactions from \SI{3.45}{GeV} for scattering on nucleons to $\sim \SI{2.5}{GeV}$ for scattering on oxygen nuclei. The effect is expected to be even more significant in larger nuclei like ${}^{40}$Ar thanks to the larger initial-state momenta.  The shift of the threshold is much larger than the typical experimental energy resolution and should therefore be clearly visible in the event spectrum, given sufficient statistics.
\end{enumerate}

\section{Summary and Conclusions}
\label{sec:Conclusion}

Accurate calculations of neutrino--nucleus interaction cross sections are pivotal for precision neutrino physics. And while controlling hadronic uncertainties within the Standard Model already presents a substantial challenge, the task becomes even more daunting when potential new interactions beyond the Standard Model are introduced

In this work, we have taken an important step forward by deriving, for the first time, a comprehensive set of neutrino--nucleus cross sections that incorporate all conceivable hadronic currents within the framework of Effective Field Theory.  We have carefully determined the relevant vector, axial vector, scalar, pseudoscalar, and tensor form factors of the nucleon. Furthermore, we have established the corresponding nucleon matrix elements and integrated them with the nuclear spectral function to account for nuclear effects. An emphasis was placed on understanding systematic uncertainties, particularly those tied to the nucleon axial form factor.

Our findings show that specific Lorentz structures, especially pseudoscalar and tensor interactions, exhibit cross sections notably enhanced compared to those of the Standard Model.  Nonetheless, there exists a considerable margin of uncertainty.  Lattice QCD predicts the axial form factor, and the pseudoscalar one which is derived from it, to be larger by about 10\% compared to the simple dipole parameterization and the $z$-expansion approach based on neutrino--deuteron scattering data. We also observed that nuclear effects hold significance across all Lorentz structures, even at multi-GeV energies. For charged current $\nu_\tau$ interactions, factoring in the initial state momentum of the impacted nucleon smears out the energy threshold.

Our results can be used in future studies of new physics in the neutrino sector to relate observed event rates to the coupling constants of an underlying theoretical model.  Further recommendations on how to use our results in practice are given in \cref{app:OscExp}.

Directions for future work include applying our methods to different target nuclei, as well as updating form factors and spectral functions with new data from neutrino--nucleus and electron--nucleus scattering experiments. Attention must be paid to the fact that this data itself may be contaminated by physics beyond the Standard Model.

\begin{acknowledgments}
It is a pleasure to thank Salvador Urrea and Noah Steinberg for very useful discussions and for collaboration during parts of this project. This manuscript was managed by Fermi Research Alliance, LLC under Contract No.\ DE-AC02- 07CH11359 with the U.S.\ Department of Energy, Office of Science, Office of High Energy Physics and by the NeuCol SciDAC-5 program (NR). The work of ZT is supported by the Neutrino Theory Network Program Grant \#DE-AC02-07CHI11359 and the US Department of Energy under award \#DE-SC0020250. She also appreciates the hospitality of Aspen Physics Center where this work was partially developed. JK would like to acknowledge support from a Fermilab Neutrino Physics Center Fellowship.
\end{acknowledgments}

\appendix
\section{Impact of New Physics on Standard Model Parameters}
\label{sec:WEFT-redefinition}

As explained in \cref{subsec:WEFT} and in refs.~\cite{Bhattacharya:2011qm, Gonzalez-Alonso:2016etj, Falkowski:2017pss, Descotes-Genon:2018foz}, the measured values of the SM parameters entering our calculations, in particular $v$ and $V_{ud}$ may themselves be affected by new physics. The resulting biases can, however, be absorbed into a redefinition of $\epsilon_L$ as we now show explicitly. The true values of the SM parameters (which we denote with a superscript `0') can formally be related to the measured values (for which we use the superscript `PDG') via
\begin{align}
    \begin{split}
        v^\text{PDG}      &\equiv v^0      + \delta v \,, \\
        V_{ud}^\text{PDG} &\equiv V_{ud}^0 + \delta V_{ud} \,.
    \end{split}
    \label{eq:v-Vud}
\end{align}
Up to first order in the new physics parameters, $\delta v$ and $\delta V_{ud}$ can be absorbed into a redefinition of $\epsilon_L$ according to
\begin{align}
    \epsilon_L \to \bar\epsilon_L \equiv \epsilon_L + 2 \delta v - \delta V_{ud} \,,
\end{align}
as can be easily seen by directly plugging \cref{eq:v-Vud} into the WEFT Lagrangian (cf.\
\cref{eq:EFT_lweft})
\begin{align} 
    \cL_{\rm WEFT} 
    &\supset
    - \,\frac{2 V_{ud}^0}{(v^0)^2} \Big\{
      [ {\bf 1} + \epsilon_L]_{\alpha\beta}
             (\bar{q}_u \gamma^\mu P_L q_d) (\bar\ell_\alpha \gamma_\mu P_L \nu_\beta)
    \,+ \ldots \Big\}
  \label{eq:EFT_lweft-bare}
\end{align}
Here, $\ldots$ stands for all new physics terms with Lorentz structures $R$, $S$, $P$, $T$. We see that, interpreting $v$ as $v^\text{PDG}$, $V_{ud}$ as $V_{ud}^\text{PDG}$, and $\epsilon_L$ as $\bar\epsilon_L$ in \cref{eq:EFT_lweft}, the effects of new physics on the Higgs vev and on the CKM matrix elements are absorbed.

\section{Complete Expressions for the Squared Amplitudes}
\label{app:AmpSq}

In this appendix, we present the full expression for the squared amplitudes, without the approximation of nucleons at rest that was used in \crefrange{eq:B-LL}{eq:D-LT}. With the definitions $D_0 \equiv 2 V_{ud}^2 / v^4$ and $\mathcal{P}_{ij} \equiv p_i \cdot p_j$, we find 
\begin{align}
\cline{1-3}
%
   \label{eq:AsqL&R} 
   \frac{1}{2} \sum_\text{spin} |A_{L(R),\alpha}|^2
        &= D_0 \Bigg\{\Big(\mathcal{P}_{\ell_\alpha n}\mathcal{P}_{\nu p}+\mathcal{P}_{\ell_\alpha p}\mathcal{P}_{\nu n}\Big) 
    \bigg[\big(2F_1+F_2\big)^2+\big(\tilde{G}_P-2G_A\big)^2+\frac{\mathcal{P}_{n p}}{M_N^2}\big(F_2^2-\tilde{G}_P^2\big)
    \bigg]\nonumber\\
        &\hspace{0.7cm}
        -\Big(\mathcal{P}_{\ell_\alpha n}\mathcal{P}_{\nu n}+\mathcal{P}_{\ell_\alpha p}\mathcal{P}_{\nu p}\Big) 
     \bigg[4F_1F_2+3F_2^2+\tilde{G}_P\big(\tilde{G}_P-4G_A\big)-\frac{\mathcal{P}_{n p}}{M_N^2}\big(F_2^2+\tilde{G}_P^2\big)
    \bigg]
        \nonumber\\
        &\hspace{0.5cm}
        \mp 8\Big(\mathcal{P}_{\ell_\alpha n}\mathcal{P}_{\nu p}-\mathcal{P}_{\ell_\alpha p}\mathcal{P}_{\nu n}\Big) 
     \bigg[G_A \big(F_1+F_2\big)
    \bigg]\\
    &\hspace{0.7cm}
    -M_N^2\mathcal{P}_{\ell_\alpha \nu} \bigg[\Big(1-\frac{\mathcal{P}_{n p}}{M_N^2}\Big)\Big(4F_1 F_2+F_2^2+\tilde{G}_P\big(4G_A-\tilde{G}_P\big)-\frac{\mathcal{P}_{n p}}{M_N^2}\big(F_2^2-\tilde{G}_P^2\big) \Big)
    \nonumber\\
    &\hspace{10cm}
    +4\big(F_1^2-G_A^2\big)
    \bigg] \Bigg\} \,,
    \nonumber\\
\cline{1-3}
%
 \frac{1}{2} \sum_\text{spin} |A_{S,\alpha}|^2
    &=2 D_0 G_S^2 \mathcal{P}_{\ell_\alpha \nu}  \Big(\mathcal{P}_{np}+M_N^2 \Big)\,,\\
\cline{1-3}
%
 \frac{1}{2} \sum_\text{spin} |A_{P,\alpha}|^2
    &=2 D_0 G_P^2  \mathcal{P}_{\ell_\alpha \nu}  \Big(\mathcal{P}_{np}-M_N^2 \Big)\,,\\
\cline{1-3}
%
 \frac{1}{2} \sum_\text{spin} |A_{T,\alpha}|^2
     &= 4D_0 \Bigg\{\Big(\mathcal{P}_{\ell_\alpha n}\mathcal{P}_{\nu n}+\mathcal{P}_{\ell_\alpha p}\mathcal{P}_{\nu p}\Big) 
     \bigg[
     2 (G_T^{(1)}+G_T^{(2)}) \Big(G_T-G_T^{(1)}-G_T^{(2)}\Big)
      \nonumber\\
     &\hspace{7cm}
      +\frac{\mathcal{P}_{n p}}{M_N^2}  \Big({G_T^{(1)}}^2-2 {G_T^{(2)}}^2\Big)
     \bigg]\\
     &\qquad
     +\Big(\mathcal{P}_{\ell_\alpha n}\mathcal{P}_{\nu p}+\mathcal{P}_{\ell_\alpha p}\mathcal{P}_{\nu n}\Big) 
     \bigg[
     2 \frac{\mathcal{P}_{n p}}{M_N^2} G_T^{(2)} \bigg(-G_T + 2 G_T^{(1)}
    + G_T^{(2)}\Big(1+\frac{\mathcal{P}_{n p}}{M_N^2}\Big)\bigg)
      \nonumber\\
     &\hspace{8.3cm}
      +\big(G_T - G_T^{(1)}\big)^2
     \bigg]
       \nonumber \\
    &\qquad
    +M_N^2 \mathcal{P}_{\ell_\alpha \nu} 
   \bigg[G_T^2  \frac{\mathcal{P}_{n p}}{M_N^2}+2\Big(1-\frac{\mathcal{P}_{n p}}{M_N^2}\Big)\bigg(G_T^{(1)}
    + G_T^{(2)}\Big(1+\frac{\mathcal{P}_{n p}}{M_N^2}\Big)\bigg)
       \nonumber\\ 
    &\hspace{6cm}
    \times \bigg(G_T-G_T^{(1)}
    - G_T^{(2)}\Big(1+\frac{\mathcal{P}_{n p}}{M_N^2}\Big)\bigg) \bigg]
      \Bigg\}\,,
      \nonumber \\
\cline{1-3} \notag
\end{align}
where in \cref{eq:AsqL&R} the $-$ and $+$ signs correspond to $|A_{L,\alpha}|^2$ and $|A_{R,\alpha}|^2$, respectively. For the interference terms with the SM we find 
\begin{align}
\cline{1-3}
%
 %
   \frac{1}{2} \sum_\text{spin} A_{L,\alpha} A^*_{R,\alpha}
        &= D_0 \Bigg\{\Big(\mathcal{P}_{\ell_\alpha n}\mathcal{P}_{\nu p}+\mathcal{P}_{\ell_\alpha p}\mathcal{P}_{\nu n}\Big) 
    \bigg[\big(2F_1+F_2\big)^2-\big(\tilde{G}_P-2G_A\big)^2+\frac{\mathcal{P}_{n p}}{M_N^2}\big(F_2^2+\tilde{G}_P^2\big)
    \bigg]\nonumber\\
        &\quad
        -\Big(\mathcal{P}_{\ell_\alpha n}\mathcal{P}_{\nu n}+\mathcal{P}_{\ell_\alpha p}\mathcal{P}_{\nu p}\Big) 
     \bigg[4F_1F_2+3F_2^2-\tilde{G}_P\big(\tilde{G}_P-4G_A\big)-\frac{\mathcal{P}_{n p}}{M_N^2}\big(F_2^2-\tilde{G}_P^2\big)
    \bigg]
        \nonumber\\
    &\quad
    +M_N^2\mathcal{P}_{\ell_\alpha \nu} \bigg[\Big(1-\frac{\mathcal{P}_{n p}}{M_N^2}\Big)\Big(4F_1 F_2+F_2^2-\tilde{G}_P\big(4G_A-\tilde{G}_P\big)-\frac{\mathcal{P}_{n p}}{M_N^2}\big(F_2^2+\tilde{G}_P^2\big) \Big)
    \nonumber\\
    &\hspace{9cm}
    +4\big(F_1^2+G_A^2\big)
    \bigg]
     \Bigg\} \,, \\
\cline{1-3}
%
   \frac{1}{2} \sum_\text{spin} A_{L,\alpha} A^*_{S,\alpha}
        &=    D_0 m_{\ell_\alpha}M_NG_S  \Big(\mathcal{P}_{\nu n}+\mathcal{P}_{\nu p}\Big)   
        \bigg[ 2F_1+F_2\Big(1-\frac{\mathcal{P}_{n p}}{M_N^2}\Big)\bigg]\,,\\
\cline{1-3}
%
 \frac{1}{2} \sum_\text{spin} A_{L,\alpha} A^*_{P,\alpha}
        &=   -D_0 m_{\ell_\alpha}M_NG_P  \Big(\mathcal{P}_{\nu n}-\mathcal{P}_{\nu p}\Big)   
        \bigg[ 2G_A-\tilde{G}_P\Big(1-\frac{\mathcal{P}_{n p}}{M_N^2}\Big)\bigg] \,,   \\
\cline{1-3}
%
\frac{1}{2} \sum_\text{spin} A_{L,\alpha} A^*_{T,\alpha}
    &= -D_0 m_{\ell_\alpha}M_N  \Bigg\{
           \Big( \mathcal{P}_{\nu n} - \mathcal{P}_{\nu p} \Big) 
           \bigg[ 2 F_1 \Big(3 G_T - 4 G_T^{(1)} - 2 G_T^{(2)} \Big) \\
    &\qquad\qquad
                + F_2\Big(5G_T-6G_T^{(1)}-2G_T^{(2)}\Big) \nonumber\\[0.1cm]
    &\qquad\qquad
                + \frac{\mathcal{P}_{n p}}{M_N^2}
                  \Big( 4 F_1 \big(G_T^{(1)} - G_T^{(12}\big)
                      - F_2 \big(G_T - 6 G_T^{(1)}\big)
                      + 2 F_2 G_T^{(2)} \frac{\mathcal{P}_{n p}}{M_N^2} \Big)
            \bigg] \nonumber\\
    &\qquad
         + \Big( \mathcal{P}_{\nu n} + \mathcal{P}_{\nu p} \Big)
           \bigg[ 2 G_A \big(3 G_T - 2 G_T^{(1)} \big) - \tilde{G}_PG_T
                + \big( 4 G_A G_T^{(1)} + G_T \tilde{G}_ P\big)
                  \frac{\mathcal{P}_{n p}}{M_N^2}
       \bigg]\Bigg\}\,. \nonumber \\
\cline{1-3} \nonumber
\end{align}

\section{Application to Oscillation Experiments}
\label{app:OscExp}

In this appendix we explain how the results of this work can be used at neutrino oscillation experiments. We will follow closely the formalism introduced at Refs.~\cite{Falkowski:2019xoe,Falkowski:2019kfn}. The first important point to note is that the new interactions introduced in the Lagrangian of \cref{eq:EFT_lweft} will in general affect not only the neutrino detection cross sections, but also the production rate. Let us assume neutrinos are produced at a source $S$ in conjunction with a charged lepton $\ell_\beta$, and that the experiment can have multiply ``sources'' (e.g.\ different meson decays, $S = \pi^\pm, K^0_S, K^0_L, K^\pm$, etc., in a neutrino beam experiment). Neutrinos scatter off the target nucleon or nucleus and produce a charged lepton $\ell_\alpha$. The differential event rate in a detector a distance $L$ away from the source is~\cite{Falkowski:2021bkq}\footnote{[About NC interactions and matter effect]}:
\begin{align}
  \frac{dR_\alpha}{dE_\nu}
    &= N_T \, \sigma^\text{SM}_\alpha(E_\nu) \,
       \sum_{\beta,S} \Phi^{S,\text{SM}}_\beta(E_\nu) \, 
       \tilde{P}^S_{\beta\alpha}(E_\nu,L) \,.
  \label{eq:masterRate}
\end{align}
In this expression, $N_T$ is the number of target particles in the detector, the SM neutrino flux for each source $S$ (in the absence of new physics) is $\Phi^{S,\text{SM}}_\beta(E_\nu)$, and the SM detection cross section is $\sigma^\text{SM}_\alpha(E_\nu) = \hat\sigma_{LL,\alpha}$. The pseudo-probability $\tilde{P}^S_{\beta\alpha}(E_\nu,L)$ is the oscillation probability convoluted with terms describing the new physics in neutrino production and detection. It is given by:
{\small
\begin{align}
  \tilde P^S_{\beta\alpha} (E_\nu,L)
    &=      \sum_{n,m} e^{-i \Delta m_{nm}^2 L / (2 E_\nu)} \nnl
    &\hspace{-1.5cm}
     \times \bigg[ U_{\beta n}^* U_{\beta m}
          +\! \sum_{X,j,k} p_{XL,\beta}^{S,jk} [\epsilon_X^{jk} U]_{\beta n}^* U_{\beta m}
          +\!\!  \sum_{X,j,k} p_{XL,\beta}^{S,jk*} U_{\beta n}^* [\epsilon_X^{jk} U]_{\beta m}
          +\!\!\! \sum_{X,Y,j,k} p_{XY,\beta}^{S,jk} [\epsilon_X^{jk} U]_{\beta n}^*
                                                [\epsilon_Y^{jk} U]_{\beta m} \bigg] \nnl
    &\hspace{-1.5cm}
     \times \bigg[ U_{\alpha n} U_{\alpha m}^*
          +\! \sum_{X,r,s} d_{XL,\alpha}^{rs}  [\epsilon_X^{rs} U]_{\alpha n} U_{\alpha m}^*
          +\! \sum_{X,r,s} d_{XL,\alpha}^{rs*} U_{\alpha n} [\epsilon_X^{rs} U]^{*}_{\alpha m}
          +\! \sum_{X,Y,r,s} d_{XY,\alpha}^{rs} [\epsilon_X^{rs} U]_{\alpha n}
                                             [\epsilon_Y^{rs} U]^{*}_{\alpha m} \bigg] \,,
  \label{eq:tildeP}
\end{align}}%
where $\Delta m^2_{nm} \equiv m_{\nu_n}^2-m_{\nu_m}^2$ are the mass squared differences, $U_{\alpha m}$ are the elements of the effective PMNS matrix in matter,\footnote{It is important to keep in mind that new charged-current interactions are often accompanied by new neutral-current interactions once the EFT defined by \cref{eq:EFT_lweft} is embedded into an ultraviolet completion that respects the SM $SU(2)$ symmetry. If the new neutral-current interactions have an $L$ or $R$ Lorentz structure, they contribute to neutrino coherent forward scattering, thereby modifying the effective PMNS matrix in matter. In addition, new scalar ($S$) interactions lead to an effective modification of the neutrino mass matrix in matter, which implies modifications to both the PMNS matrix elements $U_{\alpha n}$ and to the mass-squared differences $\Delta m_{nm}^2$. Precision studies should always take into account both neutral-current and charged-current manifestations of the new physics.} and the sums run over neutrino mass eigenstates $n$, $m$, interactions $X=L,R,S,P,T$, and quark flavor indices $j$, $k$ at the source or $r$, $s$ at the detector. The impact of the new physics effects is parameterized by the production coefficients, $p_{XY,\beta}^{S,jk}$, and the detection coefficients, $d_{XY,\alpha}^{jk}$, which quantify how strongly a new interaction affects the neutrino event rate. Roughly speaking, the $p_{XY,\beta}^{S,jk}$ give the magnitude of the new physics contribution to the neutrino fluxes relative to the SM flux. The coefficients with $X \neq Y$ describe the interference between operators with different Lorentz structures, whereas the ones with $X=Y$ describe the non-interfering terms. Notably $p_{LL,\beta}^{S,jk} = 1$ as $LL$ is the Lorentz structure of SM weak interactions. Similarly, the $d_{XY,\alpha}^{jk}$ are the ratios of the (partial) detection cross sections with and without new physics. As long as only scattering on up and down quarks is considered, they are given by

\begin{align}
     d_{XY,\alpha}^{ud}=\frac{\hat\sigma_{XY,\alpha}}{\sigma_\alpha^{\rm{SM}}}\,,
\end{align}
where $\hat\sigma_{XY,\alpha}$ are the CCQE partial cross sections including new physics from \cref{sec:XSections}, and $\sigma_\alpha^{\text{SM}}$ is the total SM CCQE cross section.

For additional details on the production coefficients and the effect of new physics on various neutrino fluxes we refer the reader to Refs.~\cite{ Falkowski:2019xoe,Falkowski:2019kfn,Falkowski:2021bkq}.

\bibliographystyle{JHEP}
\bibliography{ref}

\end{document}